\documentclass[%
aip,
amsmath,amssymb,
reprint,%
]{revtex4-2}
\usepackage{graphicx}
\usepackage{dcolumn}
\usepackage{calc}
\usepackage{bm}
\usepackage[utf8]{inputenc}
\usepackage[T1]{fontenc}
\usepackage{mathptmx}
\usepackage{etoolbox}
\usepackage{xcolor}
\usepackage{soul}
\usepackage[colorlinks=true,linkcolor=blue,citecolor=blue]{hyperref}%
\usepackage{amsfonts,amsmath,amssymb,amsthm,relsize}
\usepackage{hyperref}
\def\beq{\begin{equation}}
	\def\eeq{\end{equation}}
\def\bea{\begin{eqnarray}}
	\def\eea{\end{eqnarray}}

\makeatletter
\def\@email#1#2{%
	\endgroup
	\patchcmd{\titleblock@produce}
	{\frontmatter@RRAPformat}
	{\frontmatter@RRAPformat{\produce@RRAP{*#1\href{mailto:#2}{#2}}}\frontmatter@RRAPformat}
	{}{}
}%
\makeatother

\begin{document}
	
	\preprint{AIP/123-QED}
	
	\title{Extreme rotational events in a forced-damped nonlinear pendulum}
	
	\author{Tapas Kumar Pal}
	\address{Physics and Applied Mathematics Unit, Indian Statistical Institute, Kolkata 700108, India}
	\author{Arnob Ray}
	\address{Physics and Applied Mathematics Unit, Indian Statistical Institute, Kolkata 700108, India}
	\author{Sayantan Nag Chowdhury}
	\address{Department of Environmental Science and Policy, University of California, Davis, CA 95616, USA}
	\author{Dibakar Ghosh}
	\thanks{\color{blue}Electronic mail: diba.ghosh@gmail.com}
	\address{Physics and Applied Mathematics Unit, Indian Statistical Institute, Kolkata 700108, India}
	\date{\today}

	
	\begin{abstract}
Since Galileo's time, the pendulum has evolved into one of the most exciting physical objects in mathematical modeling due to its vast range of applications for studying various oscillatory dynamics, including bifurcations and chaos, under various interests. This well-deserved focus aids in comprehending various oscillatory physical phenomena that can be reduced to the equations of the pendulum. The present article focuses on the rotational dynamics of the two-dimensional forced damped pendulum under the influence of the ac and dc torque. Interestingly, we are able to detect a range of the pendulum's length for which the angular velocity exhibits a few intermittent extreme rotational events that deviate significantly from a certain well-defined threshold. The statistics of the return intervals between these extreme rotational events are supported by our data to be spread exponentially. The numerical results show a sudden increase in the size of the chaotic attractor due to interior crisis which is the source of instability that  
is responsible for triggering large amplitude events in our system. We also notice the occurrence of phase slips with the appearance of extreme rotational events when phase difference between the instantaneous phase of the system and the externally applied ac torque is observed.
	\end{abstract}
	
	\maketitle
	
	\begin{quotation}
Natural events like droughts, earthquakes, tsunamis, floods, global pandemics, and human-made disasters like share market crashes and power blackouts are recurrent with a low probability of occurrence but with having immediate cataclysmic impacts on human society. In the literature, such recurrent and profoundly significant incidents are referred to as extreme events. From the study of the temporal dynamics of many physical systems, large-amplitude events significantly deviating from the mean state are observed occasionally, which has a qualitative similarity, recognized from the data records and statistical distribution, with those described above natural and human-made cataclysms. This similarity encourages researchers to study dynamical systems investigating those sudden, intermittent events better to understand the origin of extreme events. Our present study considers a forced damped nonlinear pendulum with ac and dc torque and identifies a sudden expansion of the chaotic attractor through the interior crisis. This sudden expansion of the chaotic attractor is connected to the origination of extreme rotational events, and our numerical simulations suggest their return interval distributions are of exponential type. System dynamics experience a phase slip during the transition from libration to rotation. Consequently, we uncover the same large phase slip during the appearance of these large-amplitude rotational events. Our research offers valuable insights into the emergence of extreme rotational events on dynamical systems and may find applicability for a better understanding of the continuous-time systems with a strange attractor.

	\end{quotation}

	\section{\label{sec:level1}Introduction}
	The study of extreme events \cite{chowdhury2022extreme,mishra2020routes,farazmand2019extreme} has the utmost importance in many scientific and interdisciplinary disciplines for their immediate severe consequences and potential applications. It is hardly possible to define, what indeed the extreme events (EEs) are, in literature. The events or phenomena with large deviation from the regular behavior having a huge impact in the society are usually contemplated to be as EEs. These recurrent EEs are observed in several natural and engineering systems. EEs present several unique challenges because they are unpredictable and occur spontaneously. EEs have received a lot of attention from experts nowadays because of their disastrous and terrible consequences on the socioeconomic situation \cite{chowdhury2022extreme}. EEs are found to occur in nature as well as it may be human-made as well. The natural events such as floods \cite{boers2014prediction}, tsunamis \cite{mascarenhas2006extreme}, earthquakes \cite{sornette1996rank}, cyclones \cite{dowdy2017extreme}, droughts \cite{nott2006extreme}, seismic activity \cite{zaccagnino2022correlation}, wildfires \cite{gomez2022fire}, volcanoes \cite{thelen2022trends}, to name but a few, and the man-made system's disasters such as power blackouts \cite{sharma2021major}, the nuclear leakage in Chernobyl and Fukushima \cite{clero2021lessons}, regime shifts in ecosystems \cite{biggs2009turning, scheffer2003catastrophic, folke2004regime}, share market crashes \cite{krause2015econophysics} are all considered as EEs. The necessity of studying EEs basically lies in restraining their adverse huge impact in terms of havoc concerning the importance of prediction \cite{cavalcante2013predictability,zamora2014suppression,amil2019machine,meiyazhagan2021model, ray2021optimized, meiyazhagan2022prediction} and mitigation \cite{cavalcante2013predictability,zamora2014suppression,suresh2018influence, ray2019intermittent, farazmand2019closed, sudharsan2021constant}. 
	

\par Generally, the events with amplitude larger than four to eight times the standard deviation from the central tendency (mean state) of the events \cite{chowdhury2022extreme} or the events whose amplitudes are in the 90th-99th percentile of the probability distribution \cite{mishra2020routes} are defined as EEs. The EEs are being occurred far away from the mean state of the skewed distribution, they appear on the tail of the distribution having less frequency of occurrence \cite{chowdhury2021extreme}. The scientific community needs help to investigate their unpredictable occurrences due to the availability of a limited amount of real data and has recently often resorted to the classical dynamical system approach \cite{chowdhury2021extreme,ansmann2013extreme}. The dynamical systems are being recognized as the prognostication to get rid of the problem of having a small number of real data \cite{chowdhury2021extreme}. Specifically, in dynamical systems, evolving the equations of motion forward in time, we may gather a huge number of simulated data which are helpful for statistical analysis \cite{chowdhury2021extreme,kaviya2020influence}. Researchers often struggle to explain the origin of EEs in natural systems. In that situation, dynamical models might facilitate the same. In the study of temporal dynamics of many dynamical systems, the occurrence of infrequent but recurrent having comparatively high or low amplitude events might have qualitative similarities with occasional large events being recorded in many real-world phenomena. The emergence of EEs are reported in several dynamical systems such as FitzHugh-Nagumo oscillators \cite{ansmann2013extreme, karnatak2014route, saha2017extreme, saha2018riddled, bialonski2015data,varshney2021traveling}, Hindmarsh-Rose model \cite{mishra2018dragon}, Li\'{e}nard system \cite{kingston2017extreme}, Ikeda map \cite{ray2019intermittent}, Josephson junctions \cite{ray2020extreme}, Ginzburg-Landau model \cite{kim2003statistics}, nonlinear Schr\"{o}dinger equation \cite{galuzio2014control}, micromechanical system \cite{kumarasamy2018extreme}, climatic models \cite{bodai2011chaotically}, ecological model \cite{sen2022influence}, mechanica system \cite{sudharsan2021symmetrical}, electronic circuits \cite{de2016local}, to name but a few. Besides, we also find some experimental evidences of appearance of EEs such as in laser systems \cite{bonatto2011deterministic}, epileptic EEG studies in rodents \cite{pisarchik2018extreme}, annular wave flume \cite{toffoli2017wind}, laser systems \cite{reinoso2013extreme}, and so on.
		
\par The emergence of EEs in dynamical systems is basically due to the presence of region of instability in the state space of the system \cite{farazmand2019extreme, sapsis2018new}. The occasional visit of a chaotic trajectory in the region of instability of the state space immediately leads to the traversing locations in the state space far away from the bounded region, after short duration the trajectory returns back to that region. As a result, the manifestation of occasional comparatively large amplitude events is observed in the temporal dynamics of the observable \cite{sapsis2018new}. The emergence of EEs in dynamical systems most of the time follows the sudden enlargement of the size of a chaotic attractor through an interior crisis which is a considerably important one among all other possible mechanisms \cite{zamora2013rogue, ray2019intermittent, PhysRevE.90.022917, ray2020understanding,kaviya2020influence,sudharsan2021emergence, kaviya2023route}. Interior crisis \cite{grebogi1987critical,grebogi1982chaotic, celso1983crises, PhysRevE.107.024216} occurs due to the collision of a chaotic attractor with the stable manifold of an unstable fixed point or an unstable periodic orbit. In multistable systems, under the presence of noise, a sudden transition from one state to another may cause the origination of EEs \cite{pisarchik2011rogue, jaimes2022multistability}. There are several other mechanisms behind the emergence of EEs in dynamical systems such as breakdown of quasiperiodicity \cite{PhysRevLett.97.210602}, intermittency \cite{kingston2017extreme, suresh2020parametric}, transition between the librational motion to rotational motion \cite{ray2020extreme, ray2022extreme}, and attractor bubbling \cite{cavalcante2013predictability,chowdhury2019synchronization,chowdhury2020distance}.
	
\par In the realm of physics and natural phenomena, the pendulum has become one of the paradigms of study. In this work, we consider a forced damped pendulum \cite{brandt2006synchronization}, and the dynamics of this system is phenomenologically rich. This system exhibits two kinds of motion, i.e., libration and rotation, as usual \cite{strogatz2018nonlinear}. Here, we investigate that extreme events may be emerge in the rotational dynamics of the damped pendulum under the inﬂuence of dc and ac torque as angular velocity becomes infrequently faster than normal during rotation for a very short period of time.  We use existing nonlinear theories to demonstrate the abrupt enlargement of the chaotic attractor through the interior crisis within a range of the pendulum's length. When the large-amplitude rotational events repeatedly exceed a certain threshold, we refer to these occurrences as ``\textit{extreme rotational events}" (EREs). Also, we depicts histogram plot that exhibits the probability of occurrences of events. This plot underpins the non-Gaussian distribution. Also, histograms of inter-arrival between extreme rotational events are plotted for two different parameter values and are closely ﬁtted by exponential distributions. 

\par The layout of this manuscript is as follows; we delineate the model's description in Sec.\ \eqref{sec:level2}. We further detail how our procedure deﬁnes EREs in Sec.\ \eqref{sec:level4}. The bifurcation analysis, time series and phase portrait plotting, statistical analysis of EREs, and subsequent results are illustrated in Sec.\ \eqref{sec:level4}. Finally, we conclude with a concise summary and future perspectives in Sec. \eqref{sec:level5}.

\section{\label{sec:level2}Model Description}

 \begin{figure}[hpt]
	\includegraphics[scale = 0.47]{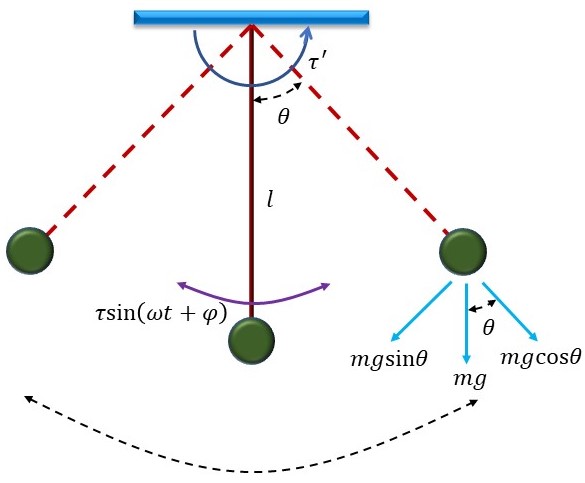}
	\caption{{\bf A schematic of a forced, damped nonlinear pendulum in the presence of dc and ac torque:} The angle between the pendulum of length $l$ and the downward vertical is denoted by the angular variable $\theta$. Here, $g$ is the acceleration due to gravity, and $m$ is the mass of the bob. Both constant dc torque $\tau'$ and periodic ac torque $\tau$ are applied to drive the pendulum counterclockwise. Here, $\omega$ is the angular frequency, and $\phi$ is the initial phase of the ac torque.}
	\label{fig1}
\end{figure}

We consider a forced damped nonlinear pendulum \cite{brandt2006synchronization} having the governing equation as 
\begin{align}
	ml^2\ddot{\theta}+\gamma \dot{\theta} = -mgl~sin\theta+\tau'+\tau~sin(\omega t + \phi)
	\label{eq.1}	
\end{align}

Here, $\theta$ is the phase variable, and $\dot{\theta}$ and $\ddot{\theta}$ denote the angular velocity and angular acceleration of the pendulum, respectively. $g$ is the acceleration due to gravity, $m$ is the mass of the bob, $l$ is the length of the pendulum, and $\gamma$ is
the damping parameter. $\omega$ is the angular frequency, $\phi$ is the initial phase of the ac torque, $\tau$ is the ac torque, and $\tau'$ is the dc torque. A schematic diagram of the pendulum \eqref {eq.1} is delineated in Fig.\ \eqref {fig1}. The angle between the downward vertical and the pendulum in this case is denoted by $\theta$. Besides, we delineate other parameters through this ﬁgure. The parameter values $m = 1.0$, $g = 1.0$, $\gamma = 0.75$, $\tau = 0.4$, $\tau' = 0.7167$, $\omega = 0.25$, and $\phi = 22/7$ are held constant throughout the text. In the following section, we examine the impact of the pendulum length $l$ by treating it as the bifurcation parameter.

	\begin{figure}[hbt]
	\centerline{\includegraphics[scale = 0.3]{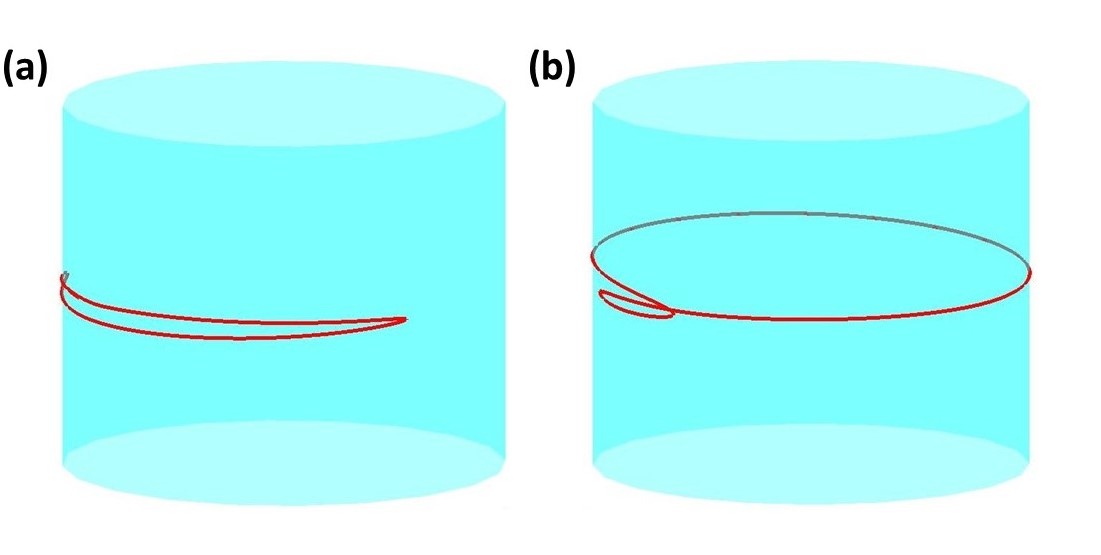}}
	\caption{ {\bf A schematic of librational and rotational orbits in cylindrical phase space:} (a) The librational orbit covers a portion of the periphery of the phase space. (b) The rotational orbit rounds the circumference of phase space.}
	\label{fig2}
\end{figure}

{\color{blue}
	
	\begin{figure*}[hpt]
			\includegraphics[scale = 0.36]{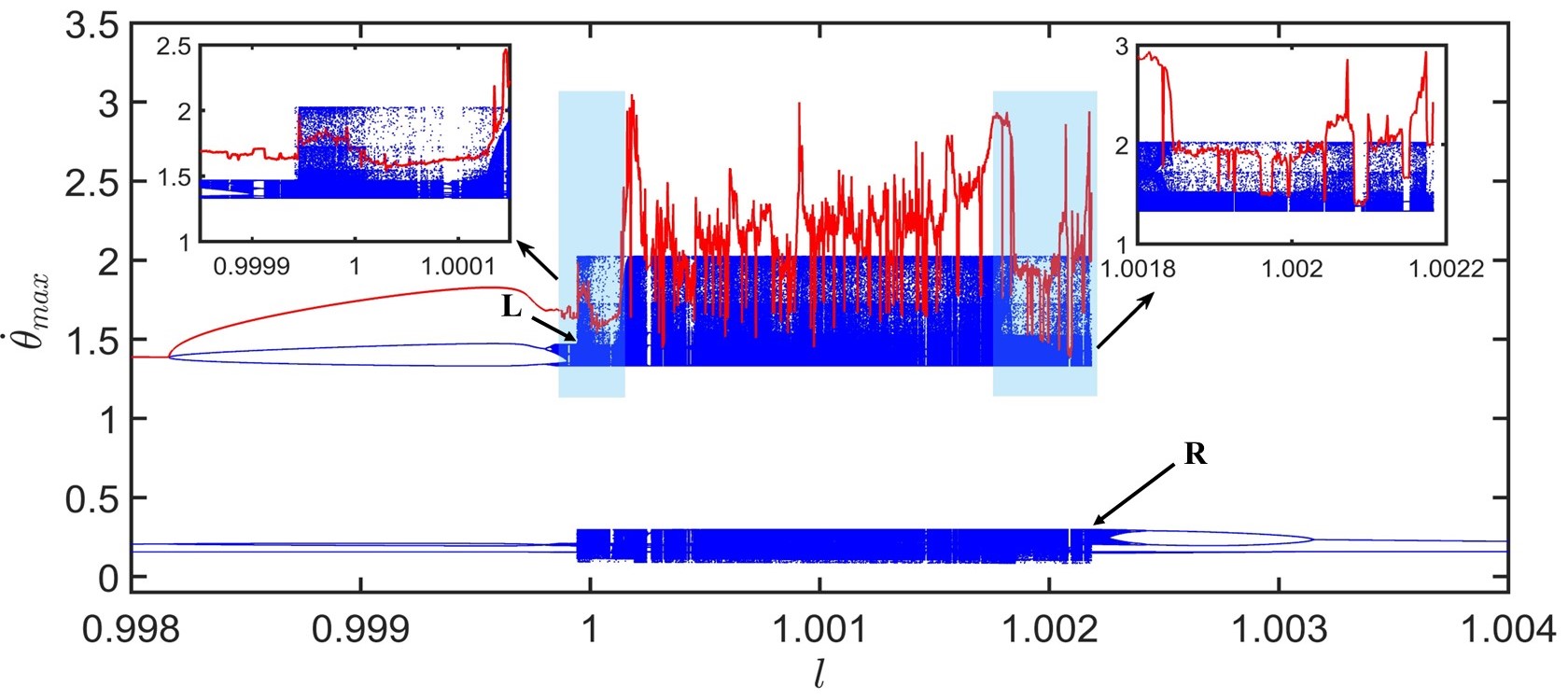}
			
		\caption{{\bf Emergence of EREs caused by the interior crises}: We draw the bifurcation diagram of $\dot\theta_{max}$ for the forced-damped nonlinear pendulum \eqref{eq.1} considering the pendulum' length $l$ as the bifurcation parameter varying in the range [0.998, 1.004] with the step length $0.00001$. Numerical simulation is performed using the RKF45 method with integration step length $0.01$ and $8 \times 10^5$ iterations, leaving a transient of $3 \times 10^5$ iterations. The pendulum displays the librational dynamics for $\dot\theta_{max} < 0.5$ and the rotational motion for $\dot\theta_{max} > 0.5$. A sudden transition from the chaotic oscillation (pre-crisis) to comparative large-amplitude chaotic oscillation (post-crisis) is observed when the value of $l$ is increased from the left-hand side of the diagram. The critical value of $l$ is indicated by {\bf L} where sudden expansion of the attractor occurs. Similarly, a sudden transition of the small amplitude chaotic oscillation in libration to large-amplitude chaotic oscillation in libration and rotation occurs at the critical value of $l$ is indicated by {\bf R} when the value of $l$ is decreased from the right-hand side of the diagram. The red line is the extreme rotational events qualifying threshold line $H_T$. Enlarge versions of the transition from two sides in rotation dynamics (two shaded portions of the bifurcation diagram) are presented in two insets on the figure's left and right sides. The left and right inset figures portray how the chaotic attractor suddenly enlarges through the interior crises. Those intermittent, sporadic blue points from the post-crisis attractor cross the red threshold line $H_T$ from the left and right sides, respectively. We set the initial condition fixed at $({\theta}_0,\dot{{\theta}_0})= (0.01,0.02)$. Although the result remains qualitatively the same for other choices of initial conditions too. Other parameter values: $\omega$=0.25, $\phi$=$\frac {22}{7}$, $m$=1.0, $g$=1.0, $\gamma$=0.75, $\tau'$=0.7167, $\tau$=0.4.} 
	\label{fig3}
\end{figure*}

}

In general, two types of motion \cite {https://doi.org/10.1002/zamm.201700007} are possible for a pendulum model. One is librational motion \cite {fistul2000libration} (small amplitude oscillation) in which the pendulum merely swings back and forth but does not fully rotate (around) with respect to the pivot, and other one is rotational motion \cite {xu2007approximate} (large amplitude oscillation) in which the pendulum fully rotates or swings around with respect to the pivot. A schematic diagram of cylindrical phase space is plotted where the trajectory of the librational orbit is shown in Fig.\ \eqref {fig2} (a). On the other hand, we observe a rotational orbit's trajectory in Fig.\ \eqref {fig2} (b).


\begin{figure}[hpt]
	\centerline{\includegraphics[scale = 0.68]{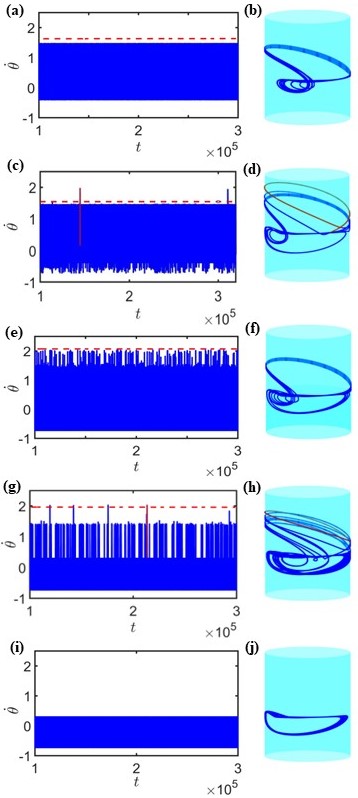}}
	\caption{{\bf Temporal dynamics (left panels) and cylindrical phase spaces (right panels)}: We choose five different values of $l$ from five different zones of the bifurcation diagram Fig.\ \eqref {fig3}. Subfigures (a) and (b) show the dynamics of the pendulum \eqref {eq.1} at $l = 0.999941$, just before the crisis point {\bf L}. Clearly, there are no signs of EREs. We plot the extreme rotational events qualifying threshold $H_T = \mu + 6\sigma$ by the red dashed horizontal lines in all subfigures of the left column. Similarly, we do not find any ERE in the third (subfigures (e-f) for $l = 1.001$) and fifth rows (subfigures (g-h) for $l = 1.002184$). We choose the value of $l$ in the third row far away from both crisis points  {\bf L} and {\bf R}. The last row only contains librational motion; hence, we do not plot the threshold $H_T$ in subfigure (i). The second (subfigures (c-d) for $l = 0.999945$) and fourth rows (subfigures (g-h) for $l = 1.00218$) reveal the appearance of EREs, as we choose the value of $l$ near the crisis points from post-crisis regimes. For illustration, two spikes (brown color) are chosen from both subfigures of time evolutions to visualize the path of trajectories corresponding to two phase spaces during extreme rotational events. All subfigures on the left column display the temporal evolutions of $\dot{\theta}$, and their respective phase spaces are shown in the right column. The values of $H_T$ are $1.634$, $1.629$, $2.189$, $1.971$ for respective values of $l=0.999941, 0.999941, 0.999941$,and $1.00218$. Other parameter values: $\omega = 0.25$, $\phi = \frac {22}{7}$, $m = 1.0$, $g = 1.0$, $\gamma = 0.75$, $\tau' = 0.7167$, $\tau = 0.4$. Initial condition: $({\theta}_0,\dot{{\theta}_0})= (0.01,0.02)$. Please see the main text for further details.}
	\label{fig4}
\end{figure}

\section{\label{sec:level4}Result}

{\it Brief overview of this section}: This section introduces the results of this study on extreme rotational events (EREs). The starting point of this work refers to how we characterize events, librational events, rotational events, and EREs. This small discussion provides the essential background to interpret our findings for dynamical system \eqref {eq.1}. The second portion of these results describes how the interior crises give rise to an enlarged attractor and crisis-induced intermittent behavior in our considered pendulum \eqref {eq.1}. All these results lie at the heart of our work. With the help of bifurcation and existing nonlinear theories, we investigate how a trajectory spends most of its time on the post-crisis attractor and occasionally does brief intermittent excursions to distant regions. Also, we explore the statistics of EREs in the last part of this section, which enables us to conclude the inter-event intervals between these EREs are distributed according to the exponential distribution, and their amplitude maintains a non-Gaussian distribution.

{\it Quantification of EREs}: Gavrielides et al.\ \cite{gavrielides1998spatiotemporal} illustrated the emergence of chaotic regimes for our chosen system described by Eq.\ \eqref{eq.1} by examining the bifurcation analysis with the variation of the length $l$ of the pendulum. They identified the approximate range of $l \in (0.998, 1.002)$ for which chaotic dynamics occur in the system. Also, it is mentioned in Ref.\ \cite{brandt2006synchronization} that the dynamics of a pendulum, given by Eq.\ \eqref {eq.1}, exhibit librational motion for $l > 1.002$, whereas it shows rotational motion for $l < 0.998$. Those studies drive us to focus on the regime of $l$, which is near the transition in dynamics from rotation to libration. In the bifurcation diagram presented in Fig.\ \eqref {fig3}, we depict the variation of the local maxima of $\theta$ by varying
$l$ and also observe the transition between high and low amplitude oscillations. Here, we can expect the occurrence of extreme events because already a few notable Refs.\ \cite{ray2020extreme, ray2022extreme} confirms the appearance of extreme events in the two-dimensional phase model during the transition between rotation and libration. So for our study, the angular velocity ($\dot{\theta}$) is the {\it observable}\ \cite{sapsis2018new} where we expect to observe the extreme events. We consider the local maxima ($\dot{\theta}_{max}$) of $\dot{\theta}$ as {\it events}. Since we know trajectory bounds only for a portion of the circumference of the cylindrical phase space due to libration, system dynamics exhibits small oscillations, and the values of $\dot{\theta}_{max}$ become lower. On the other hand, the trajectory revolves around the cylinder for rotation, resulting in large amplitude oscillation. So, the values of $\dot{\theta}_{max}$ become higher. This clear distinction is observed in the bifurcation diagram from Fig.\ \eqref {fig3}. In the present investigation, the events are classified into two classes: (a) librational events and (b) rotational events, based on this observation. We choose a threshold in such a way that large and small amplitude events are easily separated. We set the threshold value as $0.5$ since the librational and rotational events are distinguishable as a gap is observed between small and large amplitude events in the same bifurcation diagram. For $\dot{\theta}_{max}<0.5$, the events appear due to the librational motion of the system, and being so is termed as librational events. For $\dot{\theta}_{max}>0.5$, the events occurred due to rotation. We call them rotational events. In our present study, we mainly concentrate on the rotational dynamics of the pendulum \eqref{eq.1} because we observe from the bifurcation diagram that the maximum value of rotational events is, for a wide range of $l$, less than $1.5$. Still, it crosses $2$ for another range of $l$. This difference and temporal dynamics of observable lead to categorizing a subset of rotational events as EREs. To distinguish EREs from rotational events, we adopt the threshold-based statistical measure \cite{mishra2020routes, chowdhury2022extreme, ray2022extreme}, which is commonly used to classify an event as extreme events in dynamical system-related studies. A rotational event is considerable as ERE when its amplitude crosses a threshold value, $H_T = \mu + d\sigma~(d\in \mathbb{R}\setminus\{0\})$ where we chose $d = 6$ for our study. $\mu$ and $\sigma$ signify the mean and standard deviation of a collected dataset of rotational events. The choice of $d$ sets forth how far the deviation is from the mean state. It is suitably chosen for our system so that the characterization of extreme events sustains the extreme rotational events. One of the essential characteristics of EREs is the irregular occurrence in the temporal dynamics of events. The low probable occurrence of the EREs is classified depending on how the larger value of $d$ is chosen \cite {chowdhury2020distance, chowdhury2021extreme}. 

  Throughout the study, we perform the numerical simulation by integrating Eq.\ \eqref{eq.1} using the fifth-order Runge-Kutta-Fehlberg method, having an integration step length of $0.01$.

\par	{\it Generation of EREs}: A bifurcation diagram is plotted in Fig.\ \eqref {fig3} for the depiction of the changing scenario of $\dot{\theta}_{max}$ as $l$ varies within $[0.998, 1.004]$. Initially, we observe the periodic dynamics of the oscillation, and after a certain value of $l$, chaotic dynamics emerge via period-doubling bifurcation. But the amplitude of the chaotic attractor increases after crossing a particular value of $l$. In this scenario, the pendulum swings back and forth as well as whirls over the top in a chaotic fashion. After increasing the value of $l$, we only observe that the system dynamics exhibit chaotic libration beyond a specific value of $l$. That means the pendulum swings only to and fro because the combined effect of dc and ac torque is inadequate to overcome its increased rotational inertia. After that, the system undergoes from chaotic to periodic oscillation through inverse period-doubling bifurcation. We also plot the variation of $H_{T}$ by changing $l$ in Fig.\ \ref{fig3} for identifying EREs.

	\par Temporal dynamics of $\dot{\theta}$ and the corresponding phase space (${\theta}$-$\dot{\theta}$) in the cylindrical surface are displayed in Fig.\ \eqref{fig4} for five different values of $l$. In the left panel, the temporal evolutions of $\dot{\theta}$ along with threshold, $H_T$ (denoted by red dashed line) are displayed, and the respective cylindrical phase spaces ($\theta$ versus $\dot{\theta}$) are shown in the right panel for $l=0.999941, 0.999945, 1.001, 1.00218$, and $1.002184$. Figure\ \eqref{fig4} (a) depicts the temporal evolution of $\dot{\theta}$ exhibiting large amplitude oscillation for $l = 0.999941$. No large spikes or bursts are observed here; consequently, no rotational events exceed the threshold. Corresponding phase space is shown in Fig.\ \eqref {fig4} (b), where trajectory bounds within a small portion of the periphery as well as rotates the entire cylindrical surface. Occasional large spikes are observed in Fig.\ \eqref {fig4} (c), corresponding to the temporal evolution of $\dot{\theta}$ for $l = 0.999945$ because the angular velocity $\dot{\theta}$ of the pendulum occasionally increases during rotation. Here, two rotational events that cross the red threshold line $H_T$ are treated as EREs.
	Corresponding phase space is shown in Fig.\ \eqref{fig4} (d), in which trajectory is being observed in the librational (partially rounding the periphery of the cylinder) and rotational orbit (fully rounding the perimeter of the cylinder). The trajectory rotates within a bounded region during rotation but occasionally travels far away from the region, indicating the appearance of EREs. For the sake of clarity, the presence of an ERE is depicted by the brown colored spike in the temporal dynamics of $\dot{\theta}$ in Fig.\ \eqref{fig4} (c), the respective portion of trajectory is shown by the brown color in the phase space of Fig.\ \eqref {fig4} (d). Figure\ \eqref{fig4} (e) exhibits the time series of $\dot{\theta}$ and the corresponding phase space diagram is shown in Fig.\ \eqref{fig4} (f) for $l = 1.001$ where EREs are not observed anymore. Also, the trajectory's deflection from a bounded region in the phase space is absent. Figure \ref{fig4} (g) is the depiction of the temporal evolution of $\dot{\theta}$ for $l = 1.00218$ in which some intermittently large spikes are observed. Here, few rotational events exceed $H_{T}$ and are qualified as EREs. Figure\ \ref{fig4} (h) shows the respective phase space in the cylindrical surface. Here, the trajectory evolves within a portion of the circumference of the surface and also spends some time fully rounding the periphery of the cylinder. But sometimes, the trajectory traverses around the cylindrical surface far away from its regular arrival path during rotation. The temporal dynamics of $\dot{\theta}$ for $l = 1.002184$ is depicted in Fig.\ \ref{fig4} (i), where we observe only chaotic dynamics in libration.
	Since rotational dynamics are fully terminated, and the dynamics is switched over to libration, no large intermittent spikes in the time series are observed here. The denser region in Fig.\ \ref{fig4} (i) depicts the librational chaos merely. Here, the extreme rotational event qualifying red threshold line $H_T$ is not also observed because of no rotational event. We delineate the corresponding cylindrical surface depicting the trajectory in a small portion of phase space in Fig.\ \ref{fig4}(j).

	\begin{figure*}[hpt]
		\includegraphics[scale = 0.5]{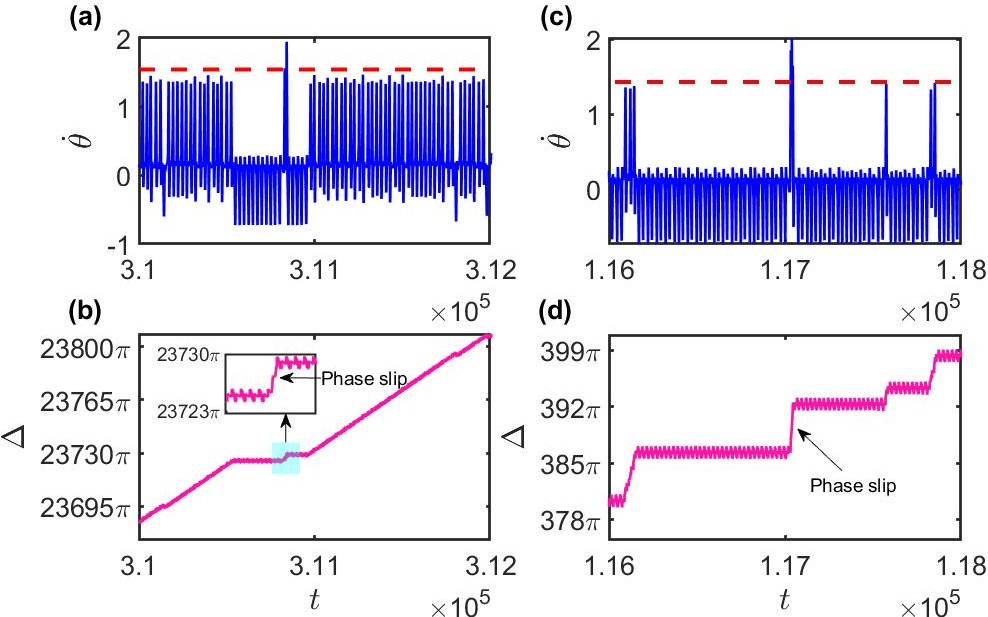}
		\caption{{\bf Time evolution of $\dot{\theta}$ (top panel) and variation of phase difference $\Delta$ with respect to time (bottom panel):} We choose two values of the pendulum's length to illustrate the temporal dynamics of $\dot{\theta}$ and the occurrence of phase slips. We choose $l = 0.999945$ for subfigures (a-b) just after the crisis point {\bf L} and $l = 1.00218$ at the crisis point {\bf R} for subfigures (c-d). The inset of subfigure (b) shows the phase slip during the manifestation of the largest spike of the time series in subfigure (a). An abrupt jump of $\Delta$ is observed during the phase slip in both subfigures (b) and (d). Red dashed lines indicate the threshold $H_T$.} 
		\label{fig7}
	\end{figure*}

	\par Now we examine the route of the emergence of extreme rotational events by analyzing the bifurcation diagram Fig.\ \eqref {fig3}. From the left side of the bifurcation diagram, a sudden and abrupt jump of the chaotic attractor is observed at the critical value of the bifurcation parameter, $l \approx 0.999942$, evinced by {\bf{L}} on the diagram when we increase the value of $l$. We also provide an enlarged version of a shaded region of the bifurcation diagram in the inset on the left-hand side for precise observation of the transition. On the left-hand side of {\bf{L}}, the dynamics are chaotic but exhibit librational as well as rotational motion, and the maximum amplitude of $\dot{\theta}$ is less than $1.5$. But, though the dynamics remain chaotic, consisting of oscillation in libration as well as rotation at the right-hand side of {\bf{L}}, the maximum amplitude of $\dot{\theta}$ reaches around $2$. Temporal dynamics of $\dot{\theta}$ near the critical value of $l$ (crisis point {\bf{L}}) is depicted in Fig.\ \ref{fig4} (c), which exhibits the appearance of EREs. Similarly, a sudden large expansion of the chaotic attractor is also noticed at the critical value of the bifurcation parameter $l \approx 1.00218$ being evinced by {\bf{R}} on the diagram from the right side of the bifurcation diagram when we decrease the value of $l$. For this scenario, chaotic librational dynamics transit to chaotic dynamics consisting of not only libration but also rotation (see the zoomed version of the shaded portion in the inset of Fig.\ \eqref {fig3} for better visualization). The temporal evolution of $\dot{\theta}$ near the critical value of $l$ is exhibited in Fig.\ \ref{fig4} (g), where the appearance of EREs is clearly detected.
	So from the left-hand side as well as the right-hand side, a sudden large expansion of the chaotic attractor occurs for the critical values of the bifurcation parameter $l$ in Fig.\ \eqref {fig3}. This incident occurs when chaotic dynamics emerge in a system through {\it interior crisis}. Chaotic attractors can experience sudden and qualitative changes depending on the system parameters. These changes are well-known in the literature as crises  \cite{grebogi1982chaotic, celso1983crises, grebogi1987critical,nag2020hidden}. A chaotic attractor, typically relatively small, collides with the stable manifold of a chaotic saddle, leading to intermittent behavior and sudden enlargement of the attractor in dynamical systems, especially during the interior crisis. This process is also responsible for the appearance of extreme events in many dynamical systems \cite{ray2019intermittent, sudharsan2021emergence, kingston2017extreme}. So, we also conclude that extreme rotation events are generated in the system through the route of interior crisis. Interior crisis-induced intermittent large spikes in time evolution of $\dot{\theta}$ are observed in Figs.\ \ref{fig4} (c) and (g). 
	
	\par Now, another interesting observation is experienced, connected with the switching between librational and rotational dynamics when the pendulum swings back and forth and whirls over the top successively. From Fig.\ \ref{fig3}, we observe oscillatory dynamics in libration and rotation until the pendulum length $l$ crosses $1.00218$. When the system switches between rotational and librational dynamics, the difference \cite{kingston2017extreme, thangavel2021extreme} between the instantaneous phases of the system and the forcing signal exceeds multiple of $\pi$. Generally, this abrupt change of phase difference is called {\it phase slip} \cite{pikovsky2002synchronization, shiju2019hilbert}. So, to verify the appearance of the phase slip, 
	we first calculate the instantaneous phase of the ac torque $\tau sin(\omega t + \phi)$ using the Hilbert Transform \cite{govindan2009understanding} method and then calculate its difference with the system's phase $\theta$, and finally, we plot the phase difference $\Delta$ with respect to time in the lower panel of Fig.\ \eqref{fig7}. Interestingly, in the case of each switching between libration and rotation, its subsequent phase slip is observed in the time evolution of $\dot{\theta}$ delineated in the upper panel of Fig.\ \eqref{fig7}. For $l = 0.999945$, the time series of $\dot{\theta}$ is depicted in Fig. \eqref{fig7} (a), and its corresponding phase difference is shown in Fig.\ \ref {fig7} (b). For the large-amplitude rotational event observed in the time series plot, its respective phase slip is shown in the inset figure for clear visualization of the abrupt jump of the phase difference $\Delta$, highlighted by the shaded portion for the specific region. In Fig.\ \eqref{fig7} (c), the time evolution of $\dot{\theta}$ for $l = 1.00218$ is portrayed, and its corresponding phase difference plot is depicted in Fig.\ \eqref{fig7} (d). The two subfigures (b) and (d) of the phase difference plot make it reasonably clear that a phase slip in the phase difference with regard to time takes place during the transition of system dynamics from libration to rotation. Consequently, phase slip occurs with the occurrence of EREs.

\begin{figure}[hpt]
\centerline{
	\includegraphics[scale = 0.7]{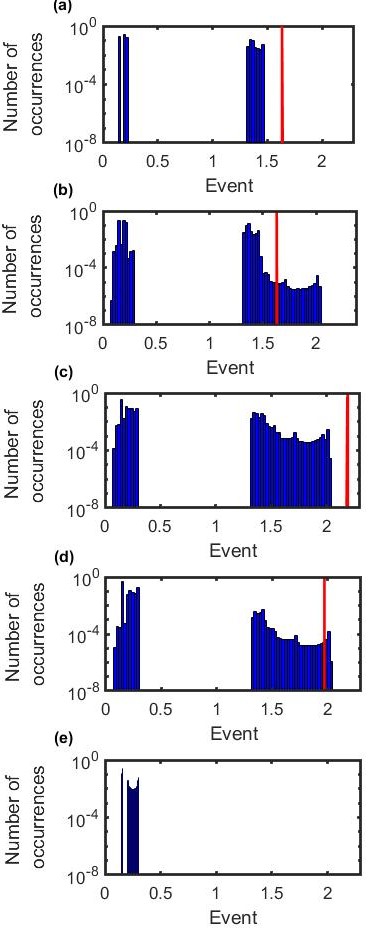}
}
\caption{{\bf Histograms of rotational and librational dynamics in semi-log scale}: We collect the data related to both librational and rotational motion over long iterations of length $10^{11}$ leaving a transient of $10^6$. The gap between the two groups is due to the presence of two different motions involving librational and rotational dynamics. The group of bins on the left side corresponds to the librational motion, and the other group on the right side has rotational dynamics. Using the red vertical line, we also plot the extreme rotational event qualifying threshold $H_T = \mu + 6\sigma$, where $\mu$ is the sample mean, and $\sigma$ is the standard deviation of the sample. This threshold helps to distinguish the EREs from the chaotic rotational dynamics. The rotational events on the left of $H_T$ are considered as EREs. Note that subfigure (e) contains only librational events at $l = 1.002184$. 
	 Parameter values: (a) $l = 0.999941$, (b) $l = 0.999945$, (c) $l = 1.001$, (d) $l = 1.00218$, and (e) $l = 1.002184$. Initial condition: $({\theta}_0,\dot{{\theta}_0})= (0.01,0.02)$.}
\label{fig5}
\end{figure}

\begin{figure}[hpt]
\centerline{
	\includegraphics[scale = 0.45]{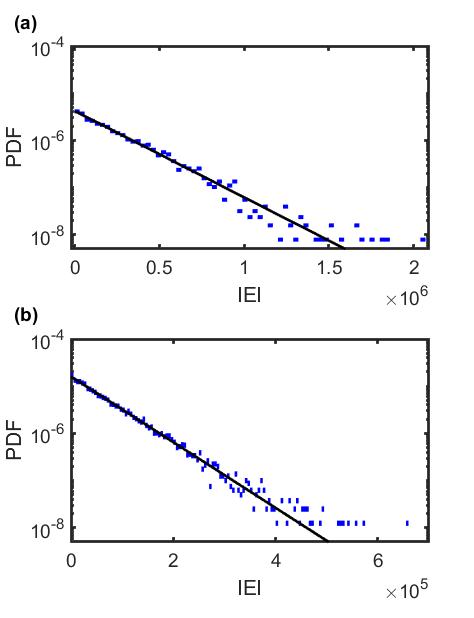}
}
\caption{{\bf {Probability density functions (PDFs) of the inter-event intervals in semi-log scale}}: We calculate the inter-event intervals (IEI) between consecutive EREs from the collected sample of size $10^{11}$ discarding a sufficiently long transient of length $10^6$. Our numerical data (blue bins) fits well with the continuous exponential distribution (black line). Small inter-event intervals have higher probabilities of occurrence of EREs. However, since the distribution is positively skewed (right-skewed), the probability of the appearance of EREs reduces significantly as the inter-event intervals increase. Parameter values: (a) $l = 0.999945$ and (b) $l = 1.00218$. Estimated rate parameter: (a) $\lambda = 4.2177 \times 10^{-06}$  with standard error $3652.39$, and (b) $\lambda = 1.6018 \times 10^{-05}$ with standard error $493.323$. Calculated coefficient of variation (CV): (a) CV = $0.9876$ and (b) CV = $1.0168$. We use the MATLAB Distribution Fitter app to fit the probability distribution to the gathered data.}
\label{fig6}
\end{figure}

\par {\it Statistics of EREs}: Still now, we observe from Fig.\ \eqref{fig3} that the whole bifurcation diagram can be classified into five regimes for $l \in [0.998, 1.004]$. Before the pre-crisis regime, i.e., for $l < 0.999942$, there is no sign of EREs. To further validate this claim, we gather sufficiently long data of length $10^{11}$, out of which we discard the initial transient of size $10^6$. We plot the histogram for $l = 0.999941$ in Fig.\ \eqref {fig5} {\bf(a)}. Clearly, there are two separate portions in these subfigures. Data related to rotational motion concentrate in the right group, whereas data related to librational motion accumulate in the left group. We further plot the red vertical threshold line $H_T$ to distinguish between extreme rotational events and chaotic oscillations. The distributed numerical data can not cross this threshold $H_T$ as $l = 0.999941$ is chosen from the pre-crisis regime. The scenario differs if we choose $l = 0.999945$ just after the crisis point {\bf L}. Here, a fair portion of the accumulated data crosses the red line as seen from Fig.\ \eqref {fig5} (b) and, thus, confirms the presence of EREs. This finding also validates our bifurcation analysis, given in Fig.\ \eqref{fig3}. We highlight a portion by the shaded box in that figure (Fig.\ \eqref{fig3}) just after the crisis point {\bf L}, from which whenever we choose a value of $l$, then we can anticipate EREs. Beyond that grey box,  those rotational motions are not high enough to cross the pre-defined threshold $H_T$ unless we choose a value of $l$ from another shaded box just after the crisis from the opposite direction. We choose a value $1.001$ of $l$ from the intermediate range far from the crisis points {\bf L} and {\bf R}. Figure \eqref {fig5} (c) shows the chaotic trajectories can not cross the threshold line $H_T$ and hence, validates our understanding. As we move further toward the point {\bf R} ($l \approx 1.00218$), we can expect the emergence of such large-amplitude EREs again. We choose a value of $l = 1.00218$ at the crisis point {\bf R}, where we expect the occurrence of EREs, as illustrated in Fig.\ \eqref{fig3}. Our grouped data into bins along the x-axis in Fig.\ \eqref {fig5} (d) attests to the occurrence of EREs. Figure \eqref{fig3} also confirms the same attribute for this choice of $l$, where we notice the bounded chaotic (rotational) trajectory explodes into a large-size attractor. This bifurcation diagram displays that there are no rotational events beyond this crisis point {\bf R}; hence, we can not anticipate such large-amplitude EREs there. Thus, we can not detect any EREs for $l = 1.002184$ in Fig.\ \eqref {fig5} (e). Figure \eqref {fig5} (e) only shows the data related to the librational motion.

\par To further study the inter-event intervals between the EREs \cite {santhanam2008return}, we choose two particular values of $l$. We select the value of $l$ initially as $0.999945$ that lies just after the crisis point ${\bf L}$. The second one, $l = 1.00218$, coincides with the crisis point ${\bf R}$. These values of $l$ correspond to the emergence of EREs, as discussed using previous figures. We plot the histograms of the gathered data of length $10^{11}$ after discarding the initial transient of size $10^6$. Throughout the study, we use the same initial conditions ${\theta}_0= 0.01$ and $\dot{{\theta}_0}=0.02$ (unless stated otherwise). We find our accumulated data are best fitted to the exponential distribution as shown in Fig.\ \eqref {fig6}. Using MATLAB, we confirm our numerical data (shown in blue bins in Fig.\ \eqref {fig6}) fitted by the following probability density function (PDF) 
	
\begin{equation}\label{eq.2}
	\begin{array}{lcl} f(x;\lambda)=
		\begin{cases} 
			\lambda e^{-\lambda x}; &  ~x \ge 0,  \\
			0; & ~x <0,
		\end{cases}
	\end{array}
\end{equation}

where $\lambda >0$ is the rate parameter. We explicitly calculate this rate parameter $\lambda$ for $l=0.999945$ and $1.00218$. The best estimated rate parameters are $\lambda = 4.2177 \times 10^{-06}$ with standard error $3652.39$ for Fig.\ \eqref {fig6} (a) and $\lambda = 1.6018 \times 10^{-05}$ is for \eqref {fig6} (b) with standard error $493.323$. Notably, the standard error in subfigure (a) is more significant (larger) than in subfigure (b) regarding data fitting by the exponential distribution. This large error is due to the number of EREs in our gathered data. We iterate the system \eqref {eq.1} for $10^{11}$ iterations discarding a sufficiently long transient of length $10^6$ for both subfigures. However, the number of EREs in subfigure (a) is $4214$, and the same in subfigure (b) is $16015$. The availability of a larger number of EREs in subfigure (b) offers a better statistical convergence to the exponential distribution than in subfigure (a). Thus, we get a lower standard error in subfigure (b). We additionally calculate the coefficient of variation (CV), defined as the ratio of standard deviation and the mean of the sample. This measure is equal to $1$ in the case of the exponential distribution. The respective CVs are $0.9876$ (for $l = 0.999945$) and $1.0168$ (for $l = 1.00218$) for the gathered sample shown in subfigures (a) and (b) of Fig.\ \eqref{fig6}. Since these derived CVs are very close to unity for our sample and using MATLAB distribution fitter, we conclude the accumulated data is distributed according to the exponential distribution.

\section{\label{sec:level5}Conclusions}
 
 \par In this work, we have shown how a suitable choice of pendulum length can produce large-amplitude rotational motion of a forced-damped nonlinear pendulum under the influence of the ac and dc torque. We have characterized the librational and rotational events using the bifurcation analysis. The same bifurcation diagram helps us detect extreme events' emergence in the rotational dynamics. The system displays rotational and librational dynamics, but occasionally the angular velocity becomes higher during rotation than in regular observation. Our numerical simulations suggest that the chaotic attractor in the rotational motion suddenly enlarges at two different post-crisis regimes due to the interior crisis, generating intermittent behavior in the rotational dynamics. The temporal evolutions of the angular velocity further validate that these sporadic rotational events occasionally cross a statistically pre-defined threshold. Hence, these large-amplitude rotational events have similar features of extreme events, as observed in various nonlinear dynamical systems. We have also displayed the respective phase portrait for each time series to confirm our claims. Furthermore, we have obtained an exponential distribution for the inter-event intervals between EREs. We have elucidated the occurrence of phase slips between the system's phase and the externally applied ac torque in due course of the origination of extreme rotational events.
 
 \par It might be interesting to detect the chaotic saddles lying in the basin of attraction of the attractor, generating interior crises. However, identifying such nonattracting chaotic sets mediating interior crises is more challenging and requires further investigation. One might investigate coupled nonlinear pendula in the spirit of the present study, which may offer greater insight into a wide range of dynamical systems. Examining whether the observed signature of extreme rotational events is experimentally observable will be further interesting. Such generalizations are left as an exciting core avenue for future research. It is also possible to investigate why the angular velocity rises irregularly during rotation.
 In conclusion, we anticipate that the findings of this work will contribute to our knowledge of how extreme large-amplitude events arise in nonlinear dynamical systems and will encourage additional research into the causes of these extreme rotational events in other non-equilibrium systems.


\section*{Acknowledgements}
DG and TKP are supported by Science and Engineering Research Board (SERB), Government of India (Project no. CRG/2021/005894). TKP is stretching his naive and earnest thanks to Gourab Kumar Sar, Md Sayeed Anwar, and S. Sudharsan for their friendly support and benevolent discussions. AR wants to thank Subrata Ghosh for fruitful discussions. SNC and AR are indebted to Arindam Mishra, Chittaranjan Hens, and Syamal K Dana for valuable discussions and feedback on this manuscript. SNC thanks Srilena Kundu for the insightful discussions and productive conversations.

\section*{Conflict of Interest}
The authors have no conflicts to disclose.


\section*{Data availability}

The data that support the findings of this study are available from the corresponding author upon reasonable request.




\label{eq.3}	

	
			
			
			
\bibliographystyle{apsrev4-1}
%

\end{document}